\documentclass[12pt,letterpaper,superscriptaddress]{JHEP3}
\usepackage{amssymb,amsmath,amsfonts,amstext,graphics, mathrsfs}
\usepackage{cite}
\usepackage{graphicx}
\usepackage{epstopdf}
\input epsf.tex

\newdimen\tableauside\tableauside=1.0ex
\newdimen\tableaurule\tableaurule=0.4pt
\newdimen\tableaustep
\def\phantomhrule#1{\hbox{\vbox to0pt{\hrule height\tableaurule width#1\vss}}}
\def\phantomvrule#1{\vbox{\hbox to0pt{\vrule width\tableaurule height#1\hss}}}
\def\sqr{\vbox{%
  \phantomhrule\tableaustep
  \hbox{\phantomvrule\tableaustep\kern\tableaustep\phantomvrule\tableaustep}%
  \hbox{\vbox{\phantomhrule\tableauside}\kern-\tableaurule}}}
\def\squares#1{\hbox{\count0=#1\noindent\loop\sqr
  \advance\count0 by-1 \ifnum\count0>0\repeat}}
\def\tableau#1{\vcenter{\offinterlineskip
  \tableaustep=\tableauside\advance\tableaustep by-\tableaurule
  \kern\normallineskip\hbox
    {\kern\normallineskip\vbox
      {\gettableau#1 0 }%
     \kern\normallineskip\kern\tableaurule}%
  \kern\normallineskip\kern\tableaurule}}
\def\gettableau#1 {\ifnum#1=0\let\next=\null\else
  \squares{#1}\let\next=\gettableau\fi\next}

\newcommand{\gsim}{\lower.7ex\hbox{$\;\stackrel{\textstyle>}{\sim}\;$}}
\newcommand{\lsim}{\lower.7ex\hbox{$\;\stackrel{\textstyle<}{\sim}\;$}}
\def\OO{{\cal O}}
\def\DD{{\cal D}}
\def\LL{{\cal L}}
\def\NN{{\cal N}}
\def\PP{{\cal P}}

\def\WW{{\cal W}}

\def\SS{{\cal S}}
\def\PP{{\cal P}}

\def\HHH{{\mathscr{ H}}}
\def\VVV{{\mathscr{ V}}}
\def\RRR{{\mathscr{ R}}}

\newcommand{\half}{{\frac{1}{2}  }}

\newcommand{\hc}{\text{ h.c. }}
\newcommand{\identity}{{\rlap{1} \hskip 1.6pt \hbox{1}}}
\newcommand{\slD}{D\hspace{-0.185in}\not\hspace{0.105in}}
\newcommand{\sld}{\partial\hspace{-0.155in}\not\hspace{0.075in}}

\newcommand{\qsl}{q\hspace{-0.142in}\not\hspace{0.10in}}

\newcommand{\bef}{\begin{figure}[htbp]\begin{center}}
\newcommand{\eef}{\end{center}\end{figure}}

\title{
\begin{flushright}
\mbox{\normalsize SLAC-PUB-13869}
\end{flushright}
\vskip 15 pt
The Simplicity of Perfect Atoms: Degeneracies in Supersymmetric Hydrogen }
\author{ Tomas Rube $^{1}$ and Jay G. Wacker $^{2,1}$\\
$^1$Stanford Institute for Theoretical Physics, Stanford University, Stanford, CA 94305\\
$^2$Theory Group, SLAC,  Menlo Park, CA 94025
}
\abstract{
Supersymmetric QED hydrogen-like bound states are remarkably similar to
non-supersymmetric hydrogen, including an accidental degeneracy of the
fine structure and which is broken by the Lamb shift.   
This article classifies the states, calculates the leading order spectrum, and
illustrates the results in several limits.
The relation to other non-relativistic bound states is explored.
}

\begin{document}

\section{Introduction}
\label{Sec: Introduction}

Supersymmetric bound states provide a laboratory for studying dynamics
in supersymmetric theories.  Bound states like hydrogen provide a framework for
understanding the qualitative dynamics of QCD mesons, a supersymmetric version
of QED can provide a qualitative picture for the symmetries and states
of superQCD mesons.  Furthermore, recent interest in dark
matter as a composite state, leads to asking how supersymmetry
acts upon these composite states \cite{Alves:2009nf, Kaplan:2009de, Lisanti:2009am, Upcoming}.

This article calculates the leading order corrections to a hydrogen-like atoms in an exactly supersymmetric version of QED.  Much of the degeneracy is broken by the fine structure and a seminal calculation was performed in \cite{Buchmuller:1981bp} for positronium, see \cite{DiVecchia:1985xm}  for an $\NN=2$ version of positronium.  Supersymmetric hydrogen is a similar except for the absence of annihilation diagrams, see \cite{Herzog:2009fw} for an independent calculation.  In the heavy proton mass limit, the supersymmetric interactions of the theory become irrelevant operators, suppressed by powers of the proton mass like the magnetic moment operator in QED and the fine structure is identical to the non-supersymmetric theory.  This article finds that fine structure spectrum of supersymmetric spectrum of hydrogen has an accidental degeneracy which is exactly analogous to the accidental degeneracy of the  $l=0$ and $l=1$ levels of the $n=2, j=\half$ state of hydrogen. The supersymmetric version of the Lamb shift lifts the residual
degeneracy and this article computes the logarithmically enhanced breaking.

The organization of the paper is as follows.  In Sec. \ref{Sec: SusyQED} the general
framework for supersymmetric QED is set up.  Particular emphasis is placed upon
the symmetries and states of the theory.  The non-relativistic limit is taken in
Sec. \ref{Sec: NR} and in Sec. \ref{Sec: HeavyP} the equivalent of the heavy proton
limit is taken.   The accidental degeneracy is demonstrated in this limit and
the organization of the states is illustrated.  Sec. \ref{Sec: FS} calculates the 
fine structure  of supersymmetric QED and the results are similar to \cite{Herzog:2009fw}.
  The Lamb shift is computed in Sec. \ref{Sec: Lamb}
and it is shown to break the residual degeneracy.  Non-Abelian Coulombic bound states are discussed in
several different circumstances in Sec. \ref{Sec: Non-Abelian Bound States} and the properties that carry over
from hydrogen are discussed.
Finally applications are discussed in Sec. \ref{Sec: Conc}. 

\section{Supersymmetric QED}
\label{Sec: SusyQED}

The field content for supersymmetric QED consists of four chiral superfield: $E$, $E^c$, $P$, $P^c$ with charges:
$-1$, $+1$, $+1$, $-1$ under an Abelian vector superfield, $V$.  The Lagrangian for supersymmetric
QED is given by
\begin{eqnarray}
\label{Eq: SusyQEDL SF}
\LL_{\text{Susy QED}} = \int\!\!d^4\theta\; K + \int\!\!d^2\theta \; W + \hc
\end{eqnarray}
where $K$ is the Kahler potential and is given by
\begin{eqnarray}
K =  \Phi^\dagger \exp(q_\phi g V) \Phi
\end{eqnarray}
where $q_\phi$ is the charge of the chiral superfield, $\Phi$, and $g$ is the gauge coupling of the $U(1)$ vector superfield.
The superpotential is given by
\begin{eqnarray}
W =  m_e E E^c + m_p P P^c + \frac{1}{4}\WW^\alpha \WW_\alpha
\end{eqnarray}
where $m_e$ and $m_p$ are electron and proton masses, respectively and $\WW_\alpha$ is the supersymmetric 
gauge field strength.
The Lagrangian Eq. \ref{Eq: SusyQEDL SF} is expanded and is given by
\begin{eqnarray}
\label{Eq: SusyQEDL Comp}
\LL_{\text{Susy QED}} = \LL_{\text{Kin.+ Mass}} + \LL_{\text{Susy Int.}}
\end{eqnarray}
where $\LL_{\text{Kin.+ Mass}}$ is the Lagrangian for the kinetic and mass terms of the theory
and  $\LL_{\text{Susy Int.}}$ contain the supersymmetric interactions.
These are
\begin{eqnarray}
\label{Eq: Free}
\nonumber
\LL_{\text{Kin.+ Mass}} &=&  \Big[\bar{e} i \slD e + \bar{e}^c + i \slD e^c +m_e( e e^c +\hc)\\
\nonumber
&&\quad+ |D_\mu \tilde{e}|^2 +|D_\mu \tilde{e}^c|^2 + m_e^2 ( |\tilde{e}|^2 + |\tilde{e}^c|^2)\Big]\\
\nonumber
&& +\Big[ \text{  terms for $P$ and $P^c$  }\Big]\\
&&- \frac{1}{4}F_{\mu\nu}^2 + \bar{\lambda} i \sld \lambda
\end{eqnarray}
where $e$ and $\tilde{e}$ denote the Weyl fermion and complex scalar from the superfield $E$
and $F_{\mu\nu}$ is the gauge field strength of the vector field and $\lambda$ is its respective
gaugino.  The gauge interactions are contained inside the covariant derivatives, $D_\mu$.
The supersymmetric interactions are 
\begin{eqnarray}
\label{Eq: Int}
\nonumber
\LL_{\text{Susy Int.}} &=&  i \sqrt{2} g  \lambda\left(  e \tilde{e}^\dagger + \cdots\right) + \hc\\
&& - \frac{g^2}{2} \left( |\tilde{e}|^2 - |\tilde{e}^c|^2 - |\tilde{p}|^2 + |\tilde{p}^c|^2\right)^2 .
\end{eqnarray}
All of the interactions of the theory are proportional to the gauge fine structure constant
\begin{eqnarray}
\alpha = \frac{g^2}{4\pi} 
\end{eqnarray}
and throughout this article, weak coupling is assumed, $\alpha \ll 1$; though not necessarily $\alpha = \alpha_{\text{EM}} = 1/137$.

This article considers the bound states of an electron (or its respective superpartners) and  a proton (and its respective superpartners).
Throughout this article, the relative masses will satisfy $m_e \le m_p$.     The remaining portion of this section  contains a description of
the symmetries and the states of the theory in Sec. \ref{Sec: Sym}, the non-relativistic limit in Sec. \ref{Sec: NR}, and the heavy
proton limit in Sec. \ref{Sec: HeavyP}.

\subsection{Symmetries and States}
\label{Sec: Sym}

The Lagrangian in Eq. \ref{Eq: SusyQEDL SF} has all of the symmetries of QED: rotational invariance and a $\mathbb{Z}_2$ spatial parity.
In addition to parity, the Lagrangian is supplemented with a $U(1)_R$ symmetry.
Parity plays a major role in the organization of the spectrum and it acts as
\begin{eqnarray}
x^\mu\rightarrow (-1)^\mu x^\mu\qquad  \theta^\alpha \leftrightarrow \bar{\theta}_{\dot{\alpha}}\qquad V \rightarrow -V\qquad   E \leftrightarrow \bar{E}^c\qquad  P \leftrightarrow \bar{P}^c
\end{eqnarray}
This implies the supersymmetric derivatives, $\DD_\alpha$ and field strength, $\WW_\alpha = \bar{\DD}^2 \DD_\alpha V$ behaves as
\begin{eqnarray}
\DD_\alpha \leftrightarrow \bar{\DD}^{\dot{\alpha}} \qquad \WW_\alpha \leftrightarrow - \bar{\WW}^{\dot{\alpha}}.
\end{eqnarray}
Most significantly,  the $U(1)_R$ symmetry {\em does not commute} with $\mathbb{Z}_{2}$ parity and are combined into an $O(2)_R$ symmetry.  The $O(2)_R$ symmetry is important because it has two dimensional irreducible representations.   The fine structure spectrum of supersymmetric hydrogen will contain
many two-fold degeneracies; however, half arise because the states are doublets of the $O(2)_R$ symmetry.

In addition to the space-time symmetries, the constituent particles are also charged under a global $U(1)_e \times U(1)_p$ flavor symmetry and the states of hydrogen are charged under the  $U(1)_{e+p}$ diagonal subgroup.   The spectrum of supersymmetric hydrogen is doubled relative to supersymmetric positronium because hydrogen is charged under $U(1)_{e+p}$.

The leading order interaction between the supersymmetric electrons and protons arise from the Coulomb interaction and the states
organize into quantum mechanical hydrogen spectrum, with state $|nlm\rangle$, given by
\begin{eqnarray}
\psi_{nlm}(x)=\langle x|n l m\rangle= R_{nl}(r) Y_{lm}(\theta,\phi) \qquad n> 0\quad l< n\qquad  |m| \le l
\end{eqnarray}
where  $R_{nl}(r)$ are the associated Laguerre polynomials and $Y_{lm}(\theta,\phi)$ are spherical harmonics with angular momentum $l$.
In addition to the bound state spectrum, there are also the continuum states that play an important role in guaranteeing a supersymmetric answer for the fine structure spectrum \cite{Buchmuller:1981bp}.  The energy spectrum at leading order is given by
\begin{eqnarray}
E^{\text{Ryd.}}_n = - \frac{ \alpha^2 \mu}{2n^2}
\end{eqnarray}
where $\mu$ is the reduced mass
\begin{eqnarray}
\mu^{-1} = m_e^{-1} + m_p^{-1} .
\end{eqnarray}
The $n^2$ fold degeneracy at each level is a result of an accidental enhanced $O(4)$ symmetry of the non-relativistic Coulomb potential.
This is broken relativistic and supersymmetric effects at $\OO(\alpha^4 \mu)$.  In addition to the spatial wave functions,  the
spin and supersymmetric portions to the wave functions, $|\SS\rangle$ and is 
\begin{eqnarray}
|\SS\rangle_{\vec{p}_{\text{cm}}} = ( E \oplus E^c{}^\dagger)_{\vec{p}=\vec{p}_{\text{cm}} + \vec{p}_{\text{rel}}/2 } \otimes (P \oplus P^c{}^\dagger)_{\vec{p}=\vec{p}_{\text{cm}}- \vec{p}_{\text{rel}}/2} .
\end{eqnarray}
Non-relativistic bound states are necessarily massive and therefore
fill out massive N=1 multiplets which are built from the Clifford vacua, $\Omega_j$,  of spin $j$.  These have the field content
of massless N=2 multiplets.  Clifford vacua are useful for classifying the excited spectra in Sec. \ref{Sec: FS}.
This tensor product can be decomposed as
\begin{eqnarray}
|\SS\rangle = \begin{cases}
\VVV & \text{Massive charged vector multiplet}\\
\HHH_2 & \text{Two massive charged hypermultiplets}
\end{cases}.
\end{eqnarray}
where $\VVV$ contains the states of a massive charged vector superfield and $\HHH_2$ contains the states of two massive charged hypermultiplets, which is equivalent to {\em four} chiral superfields.  These states are charged under the $U(1)_{e+p}$ symmetry of the theory unlike positronium.  Explicitly, the states of $\VVV$ and $\HHH_2$ are
\begin{eqnarray}
\VVV = \begin{cases}
v_\mu & \text{massive complex vector} \\
\chi_1, \bar{\chi}^c_1 & \text{massive Dirac spinor} \\
\chi_2, \bar{\chi}^c_2 & \text{massive Dirac spinor} \\
\varsigma_- & \text{massive complex scalar} 
\end{cases} 
\quad \HHH_2 =\begin{cases}
\omega_+, \omega_-  & \text{massive complex scalars}\\
\zeta_1, \bar{\zeta}_1^c & \text{massive Dirac spinor}\\
\zeta_2, \bar{\zeta}_2^c& \text{massive Dirac spinor}\\
\varpi_+, \varpi_- & \text{massive complex scalars}
\end{cases}.
\end{eqnarray}
The $O(2)_R$ symmetry forces $\varpi$, $\chi$, and $\zeta$ into doublets.
Parity acts as
\begin{eqnarray}
v_\mu \leftrightarrow  (-1)^\mu v_\mu \qquad \chi_1 \leftrightarrow \bar{\chi}^c_2 \qquad \zeta_1 \leftrightarrow \bar{\zeta}^c_2 
\end{eqnarray}
on the fields with spins 
and transforms the scalars as
\begin{eqnarray}
\varsigma_- \leftrightarrow -\varsigma_-
\qquad  
\omega_\pm \leftrightarrow \pm\omega_\pm  
\qquad 
\varpi_\pm \leftrightarrow \pm \varpi_\pm  .
\end{eqnarray}
The Coulomb potential is insensitive to the spin/supersymmetric state and 
therefore the states factorize at $\OO(\alpha^2\mu)$ as
\begin{eqnarray}
|\psi\rangle =  |nlm\rangle \otimes | \SS\rangle .
\end{eqnarray}
The tensor product between $|lm\rangle$ and $|\SS\rangle$ is performed and the states organize themselves into massive supermultiplets.  Using the notation  $\Omega_j$ to denote a massive supermultiplet built from a Clifford vacuum
with  $j$,  
{\em e.g.} $\Omega_0$ is a hypermultiplet, the the tensor product between a state with orbital angular $l$ and a multiplet built from $\Omega_j$ is given by
\begin{eqnarray}
|lm \rangle \otimes \Omega_j = \Omega_{|l-j|} \oplus \Omega_{|l-j|+1} \oplus \cdots \Omega_{l+j} .
\end{eqnarray}
The states of principle quantum number $n$ are organized into states
\begin{eqnarray}
P_n = 2\times \Omega_0 \oplus 2 \times \Omega_\half \oplus \cdots \oplus 2\times \Omega_{n -1} \oplus 1\times \Omega_{n-\half} .
\end{eqnarray}
where there is a two-fold multiplicity of states for every $\Omega_j$ except $j=n-\half$. The leading order superspin wave functions
can be calculated by acting with the susy raising operators on the Clifford vacua in the expression above.
The states where the constituents are in a $\VVV$ configuration are mixed with each other because the Clifford vacuum has non-zero spin.  There are two basis choices  
\begin{eqnarray}
\label{Eq: Distinct Wave Functions}
a^\dagger_{\alpha} \cdot \left( |l \rangle \otimes \Omega_\half\right)
\qquad \text{ vs } \qquad
| l\rangle \otimes \left( a^\dagger_\alpha  \cdot \Omega_\half \right) 
\end{eqnarray}
where $a^\dagger_\alpha$ is the supersymmetric raising operator.   The first expression
in Eq. \ref{Eq: Distinct Wave Functions} gives the proper organization of states.
This leads to mixing between states where the constituents are in the $v_\mu$ and $\varsigma_-$ configurations.
The relative admixtures can be computed with Clebsch-Gordon coefficients.

The two-fold multiplicity of integer $j$ arises from these states are doublets
of $O(2)_R$, whereas  the degeneracy for half-integer $j$ is accidental.  
For the ground state of supersymmetric hydrogen, there is no degeneracy of the  spatial wave function and therefore the states are
\begin{eqnarray}
P_1 =  2\times \Omega_0 \oplus \Omega_\half
\end{eqnarray}
where $\VVV \simeq \Omega_\half$ and $\HHH_2= 2\times \Omega_0$ .

\subsection{Non-relativistic Limit}
\label{Sec: NR}

Hydrogen is a non-relativistic bound state and it is convenient to take the non-relativistic limit of the Lagrangian in Eq. \ref{Eq: SusyQEDL Comp} in order to gain intuition about the fine structure.  
The non-relativistic limit of fermions is found by organizing the Weyl fermions of Eq. \ref{Eq: SusyQEDL Comp} into Dirac fermions
\begin{eqnarray}
\Psi_e^{\text{W}} = \left(\begin{array}{c} e\\ \bar{e}^c \end{array} \right)
\qquad
\Psi_p^{\text{W}} = \left(\begin{array}{c} p\\ \bar{p}^c \end{array} \right) .
\end{eqnarray}
After transforming from the Weyl basis where $\gamma_0$ is off diagonal to the Dirac basis where $\gamma_0$ is diagonal,
the non-relativistic limit is found by factoring out the fast $\exp(i m t)$ of the wave function and integrating out the
small components.  For instance, the non-relativistic spinor for the electron, $\psi_e$, is related to the Dirac spinor by
\begin{eqnarray}
\Psi_e^{\text{D}}\simeq e^{im_et} \left(
\begin{array}{c} 
\psi_e \\ \frac{i \vec{\sigma}\cdot \vec{\nabla}}{2 m_e} \psi_e
\end{array}
\right)
\end{eqnarray}
and the leading order Schr\"odinger Lagrangian is
\begin{eqnarray}
\LL_{e \text{ NR}} =  \bar{\psi}_ei\partial_t {\psi}_e+ \bar{\psi}_e \frac{\nabla^2}{2m_e} \psi_e  + \cdots .
\end{eqnarray}

The non-relativistic limit for the scalars is found by  first  factoring out the $\exp(i mt)$ and then by rescaling the wave functions
\begin{eqnarray}
\tilde{e} =   e^{im_et}\frac{1}{\sqrt{2 m_e}} \phi_e \qquad \tilde{e}^c = e^{i m_e t} \frac{1}{\sqrt{2 m_e}} \phi_{e^c} .
\end{eqnarray}
The non-relativistic scalars have the same  engineering dimension of $\frac{3}{2}$ as non-relativistic fermions.
The scalars also have a Schr\"odinger Lagrangian
\begin{eqnarray}
\LL_{\tilde{e} \text{ NR}} =  \phi^\dagger _e i\partial_t {\phi}_e+ \phi^\dagger_e \frac{\nabla^2}{2m_e} \phi_e  + \cdots .
\end{eqnarray}
The key aspect to non-relativistic limit of supersymmetric  QED is that the supersymmetric interactions become 
higher dimension interactions, much like the magnetic dipole of QED.  In the non-relativistic limit, the supersymmetric interactions of Eq. \ref{Eq: Int} 
become
\begin{eqnarray}
\label{Eq: Susy NR Int}
\nonumber
\LL_{\text{Susy Int}} &=&
g \bar{\lambda} \left( \frac{1}{\sqrt{m_e}}  (  \psi_e  \phi_e + \psi^*_e \phi_{e^c}) + \frac{1}{\sqrt{m_p}}  (  \psi_p  \phi_p + \psi^*_p \phi_{p^c}) \right)+\hc\\
&& \quad-\frac{g^2}{4} \left(  \frac{1}{m_e} ( |\phi_e|^2 - |\phi_{e^c}|^2) - \frac{1}{m_p} ( |\phi_p|^2 - |\phi_{p^c}|^2)
\right)^2 .
\end{eqnarray}
The leading supersymmetric corrections to the spectrum will always use one interaction involving an electron state and one involving
a proton state and therefore all of these interactions are suppressed by powers of $(m_em_p)^{-\half}$ for gaugino interactions
and $(m_e m_p)^{-1}$ for $D$-term interactions.  The next subsection considers the heavy proton limit as a simplified example  to gain intuition about the full calculation.

\subsection{Heavy Constituent Limit}
\label{Sec: HeavyP}

This section considers the limit where  the constituents become much heavier than the other
\begin{eqnarray}
\frac{m_e}{m_p} \rightarrow 0 \qquad \Rightarrow \qquad \mu = m_e .
\end{eqnarray}
In this limit the supersymmetric interactions do not contribute to the fine structure of the supersymmetric hydrogen atom.  This is seen from Eq. \ref{Eq: Susy NR Int} where all the interactions between an electron state and proton state
are suppressed by factors of $(m_e m_p)^{-n}$, $n\ge \half$.   Sec. \ref{Sec: FS} computes that fine structure
and shows that  $n=1$ because gaugino exchange only contributes at second order in perturbation theory due to parity.  There is no way of obtaining any positive powers of $m_p$ in the numerator and this immediately implies that fine structure  of supersymmetric hydrogen  given by the purely QED corrections to the Coulomb potential.  These are exactly solved for \cite{Itzykson:1980rh} and are 
\begin{eqnarray}
\label{Eq: HydrogenSpectrum}
E^{\text{Hyd}}_{nj} =\frac{  m_e}{\sqrt{ 1 +\big(\frac{ \alpha}{n - (j+\half) + \sqrt{(j+\half)^2 - \alpha^2}}\big)^2}} 
= m_e - \frac{ \alpha^2 m_e}{2n^2} + \frac{ \alpha^4m_e}{2n^3}\left(   \frac{3}{4n}-\frac{1}{j+\half}\right)
\end{eqnarray}
where $j= l$ for scalars and $j= \half \otimes l = l-\half, l +\half$ for fermions.
Sates with scalar electrons are not degenerate with the states with fermionic electrons.  
The difference arises from the Darwin term which doesn't exist for scalars as Zitterbewegung is
a property of Dirac fermions.    

This spectrum immediately implies how the wave functions of the composite supermultiplets are
organized for mass eigenstates: the light state is either a scalar {\em or} a fermion and the different states of the 
supermultiplet are created by toggling the spin of the heavier state.    For instance, the ground state for the supersymmetric
hydrogen atom  with an infinitely massive proton is given by
\begin{eqnarray}
\HHH_2 = (\psi_p\oplus \phi_p\oplus \phi_{p^c}^*)\otimes ( \phi_e\oplus \phi_{e^c}^*)\qquad
\VVV= (\psi_p\oplus \phi_p \oplus\phi_{p^c}^*)\otimes ( \psi_e)
\end{eqnarray}
and the mass splitting is given by
\begin{eqnarray}
E_{1 \Omega_\half}-E_{1\Omega_0}  =   E^{\text{Hyd}}_{1\half}-E^{\text{Hyd}}_{10} =  \frac{\alpha^4 m_e}{2} 
\end{eqnarray}
where $E_{n\Omega_j}$ denotes the energy eigenvalues of the $n$th principle state built from the Clifford vacuum $\Omega_j$ using the correct basis Eq.~\ref{Eq: Distinct Wave Functions}.
Each of the $\Omega_\half$  multiplets have the same energy eigenvalues, which is dictated by the $O(2)_R$ symmetry.

The $n=2$ level has an accident degeneracy which is exactly analogous to the $l=0$ and $l=1$ degeneracy of the Dirac at the $n=2$ level.   One of the  $\Omega_1$ states arises from the electrons in an $l=0$ spatial configuration, while the
other arises from the electrons in an $l=1$ spatial configuration.  These states have different parity and there is no symmetry
that interchanges the two.    In the normal hydrogen atom, this degeneracy is broken by the Lamb shift.  In the supersymmetric
hydrogen atom there is a possibility that the degeneracy is broken by supersymmetric interactions.  
In this case, for the supersymmetric interactions to break the degeneracy, there would be a supersymmetric hyperfine structure scaling as
\begin{eqnarray}
E_{2 \Omega_\half^{l=0}} - E_{2\Omega_\half^{l=1}} \sim \frac{ \alpha^4 m_e^2}{m_p} .
\end{eqnarray}
This article finds that this is not the case and that the degeneracy exists even with a finite proton mass
and this accidental splitting is broken by the supersymmetric Lamb shift
and the calculation is presented in Sec. \ref{Sec: Lamb}.

\subsection{Finite Mass Wave Superspin Functions}
\label{Sec: Finite M}

The superspin wave functions $|\SS\rangle$ are decomposed in terms of the elementary constituents to leading order by demanding that the multiplets $\VVV$ and $\HHH_2$ close under supersymmetry transformations.  The supersymmetry transformations are approximated on-shell,  in the non-relativistic limit for the electron states as  
\begin{eqnarray}
\hspace{-0.3in} \delta_\xi \phi_e = \sqrt{m_e} \xi_\alpha \psi_e^\alpha, 
\quad 
\delta_\xi \phi^*_{e^c} =  \sqrt{m_e}  \bar{\xi}_{\dot{\alpha}} \delta^{\dot{\alpha} \alpha}\psi_{e\alpha},
\quad 
\delta_\xi \psi_{e\alpha} = \half \sqrt{m_e} ( \bar{\xi}_{\dot{\alpha}} \epsilon^{\dot{\alpha} }{}_\alpha\phi_e + \xi_\alpha  \phi_{e^c}^*),
\end{eqnarray}
where $\sigma^0_{\alpha\dot{\alpha}} = \delta_{\alpha\dot{\alpha}}$ and $\epsilon^{\dot{\alpha}}{}_{\alpha} = \delta^{\dot{\alpha}\beta} \epsilon_{\beta \alpha }$.  The transformation for protons are
the similar.

The superspin wave functions for a finite mass proton are computed in the 
product  superfields: $P E, P^c{}^\dagger, E^c{}^\dagger, P E^c{}^\dagger, P^c{}^\dagger E$.
After evaluating these product superfields on-shell, all of the independent degrees of freedom
reside in the last superfield: $P^c{}^\dagger E$.  The superspin configurations for $\HHH_2$ and $\VVV$
are calculated by acting with the superspace projection operators $\PP_1, \PP_2, \PP_T$ that
project on to chiral, anti-chiral, and vector superfields, respectively.   
Evaluating these superfields in the non-relativistic limit gives 
%
%After redefining the non-relativistic scalars  so that they are parity eigenstates, rather than states of $U(1)_R$
%%
%\begin{eqnarray}
%\phi_{e \pm } = \frac{1}{\sqrt{2}}( \phi_{e} \pm \phi^*_{e^c})
%\qquad 
%\phi_{p \pm } = \frac{1}{\sqrt{2}}( \phi_{p} \pm \phi^*_{p^c}).
%\end{eqnarray}
%%
%Supersymmetry acting upon the composite states
%%
%\begin{eqnarray}
%|\SS\rangle = ( \phi_{e+} \oplus \phi_{e-} \oplus \psi_e) \otimes ( \phi_{p+} \oplus \phi_{p-} \oplus \psi_p)
%\end{eqnarray}
%%
%closes into representations 
%%
\begin{eqnarray}
\VVV &=& \begin{cases}
v_\mu   & \vec{v} = \psi_p\vec{\sigma}\psi_e \ \ \\
\chi_1, \bar{\chi}^c_1&\psi_{\chi_1}=c_\theta \phi_{p} \psi_e - s_\theta \psi_{p} \phi_{e}\\
\chi_2, \bar{\chi}^c_2 &\psi_{\chi_2}= c_\theta\phi_{p^c}^*\psi_e-s_\theta\psi_p \phi_{e^c}^* \\
\varsigma_- &\varsigma_-= c_{2\theta} \psi_p \psi_e - s_{2\theta} (\phi_{p}\phi_{e^c}^* - \phi_{p^c}^*\phi_{e})/\sqrt{2}
\end{cases} \\
\ \HHH_2 &=&\begin{cases}
\omega_+ & \omega_+ = (\phi_{p} \phi_{e^c}^* + \phi_{p^c}^*\phi_{e})/\sqrt{2} \\
\omega_-& \omega_-= c_{2\theta} (\phi_{p}\phi_{e^c}^* - \phi_{p^c}^*\phi_{e})/\sqrt{2}+s_{2\theta} \psi_p \psi_e \\
\zeta_1, \bar{\zeta}_1^c & \psi_{\zeta_1}=c_\theta \psi_p \phi_{e} + s_\theta \phi_{p} \psi_e\\
\zeta_2, \bar{\zeta}_2^c&\psi_{\zeta_2}= c_\theta \psi_p \phi_{e^c}^* + s_\theta \phi_{p^c}^* \psi_e\\\
\varpi_+ & \varpi_{+}= (\phi_{p} \phi_{e} + \phi_{p^c}^*\phi_{e^c}^*)/\sqrt{2}\\
\varpi_-& \varpi_{-}=  (\phi_{p} \phi_{e} - \phi_{p^c}^*\phi_{e^c}^*)/\sqrt{2}
\end{cases}
\end{eqnarray}
where the righthand equations label the on-shell degrees of freedom and the mixing angle is defined as
\begin{eqnarray}
\tan \theta = \sqrt{ \frac{ m_e}{m_p}} .
\end{eqnarray}
The $\theta\rightarrow 0$ limit of these  superspin wave functions demonstrates the separation of
the valence particles into the spin $\half$ electrons in $\VVV$ and the spin $0$ electrons in $\HHH_2$
described in Sec. \ref{Sec: HeavyP}.

\section{Fine Structure in Supersymmetric Hydrogen}
\label{Sec: FS}

This section presents the fine structure of supersymmetric hydrogen.  Therefore, the full
analysis will not be presented in this article and \cite{Herzog:2009fw} provides a detailed 
description.  Instead, a simplified calculation is presented that will illustrate the organization
of the states and only uses first order perturbation theory.

The standard method to calculate the energy spectrum to bound states starting from a field theoretic, effective Lagrangian is to use Breit's formulation of the effective Hamiltonian \cite{LandauLifschitz}.   
For instance, in QED the one photon exchange gives rise to a Born amplitude of
\begin{eqnarray}
\nonumber
\mathcal{T}(p_1, p_2; p_3,p_4) &=&  e^2 (\bar{u}_3 \gamma^\mu u_1) D_{\mu\nu} (\bar{u}_4 \gamma^\nu u_2)\\
&=&- \sqrt{ (2 E_1)(2E_2) (2 E_3)(2E_4)}  \;\Big[ w^{\dagger}_3 w^{\dagger}_4\;\;  U(\vec{p}_1, \vec{p}_2, \vec{q})\;\;w_1 w_2 \Big]
\end{eqnarray}
where $U(\vec{p}_1, \vec{p}_2, \vec{q})$ is the Fourier transform of the interaction Hamiltonian and $w$ are the two-component spinor  plane-wave solutions  
to the Dirac equation which is generally spin-dependent.
The coordinate space representation, $\vec{r}$, is the Fourier transform of $\vec{q}$  in the center of mass frame, $\vec{p}=\vec{p_1}=-\vec{p_2}$.  After including the relativistic corrections to energy and the higher order terms from the Born amplitude, there is a common  correction to all constituent configurations given by
\begin{eqnarray}
\Delta H_{
\text{common}}=-\left(\frac{1}{m_p^3}+\frac{1}{m_e^3}\right)\frac{\vec{p}^4}{8}-\frac{\alpha }{2m_pm_e}\left(\frac{\vec{r}\cdot(\vec{r}\cdot\vec{p})\cdot\vec{ p }}{r^3}+\frac{\vec{p}^2}{r} \right) .
\end{eqnarray}
The primary challenge of computing the leading order correction to the fine structure is that 
it is necessary to use both first and second order perturbation theory.  The need for second order perturbation
theory arises from the fact that the matrix element for gaugino
exchange between two bound states is parametrically given by
\begin{eqnarray}
\Delta E_\psi^{(1) \text{ gaugino}} = \langle \psi | V_{\text{gaugino}}|\psi\rangle= \frac{\alpha}{ \sqrt{m_e m_p}} \Big\langle \psi \Big| \int\!\!d^3q  \frac{e^{iq r}}{\qsl} \Big|\psi \Big \rangle \simeq \frac{\alpha}{\sqrt{m_em_p}}\alpha^2 m_e^2
\end{eqnarray}
where the non-relativistic limit of gaugino interaction is described in Sec. \ref{Sec: NR}.  Due to parity, there is no leading order contribution from this
interaction\footnote{Dispersive contributions from $q^0\simeq  \vec{q}^2/m_p$ are further suppressed.}.  However, at second order
\begin{eqnarray}
\Delta E_\psi^{(2) \text{ gaugino}} =  \sum_{\psi'} \frac{ | \langle \psi| V_{\text{gaugino}}|\psi'\rangle|^2}{E_\psi - E_{\psi'}}  \simeq \frac{ \alpha^4 m_e^2}{ m_p} .
\end{eqnarray}
These second order effects cancel off terms from first order perturbation theory that would have led to supersymmetry breaking hyperfine interactions.
Furthermore, the full spectrum of the hydrogen atom contributes (including the continuum states) to the second order correction.  
Schwinger's exact $O(4)$ symmetric Green's functions were used in \cite{Buchmuller:1981bp} to get an exact,  supersymmetric answer.  

Since the dynamics are supersymmetric and the final result must be supersymmetric, there a is way to side-step the second order calculation and  get the exact answer from first order perturbation theory.  In each of the multiplets of a given energy level, there is a
state that receives  no second order contribution to the energy eigenvalues.  Therefore, the first order calculation is the final $\OO(\alpha^4)$ answer.

\begin{figure}[htbp]
\begin{center}
\includegraphics[width=3in]{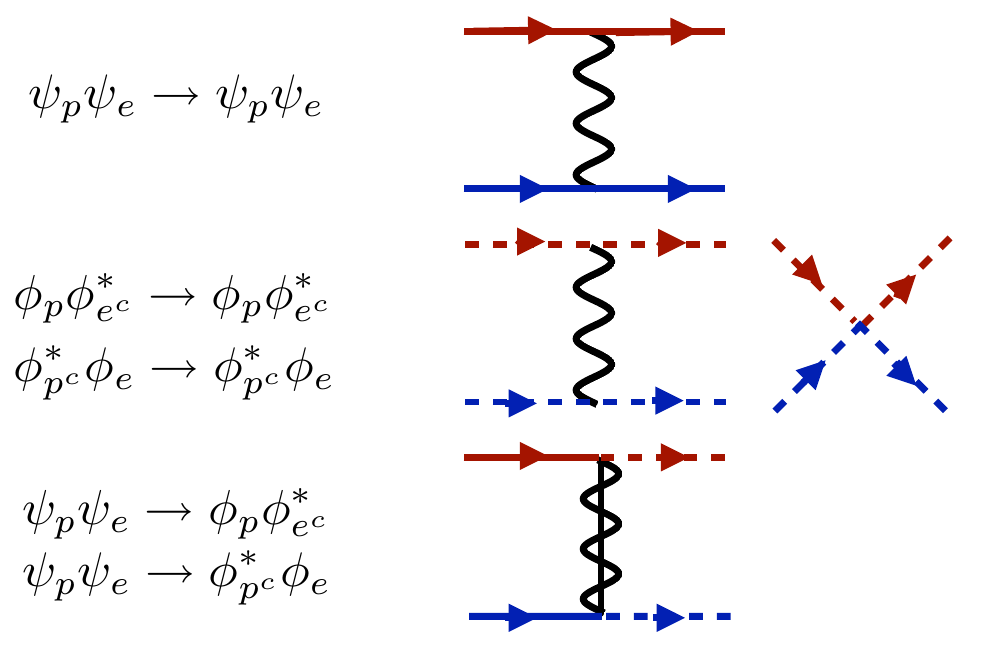}
\caption{\label{Fig: Interactions} The relevant interactions that contribute to the interaction Hamiltonian for the first-order exact states:
$\vec{v}$, $\varsigma_-$, $\omega_-$}
\end{center}
\end{figure}

\begin{figure}[htbp]
\begin{center}
\includegraphics[width=5.5in]{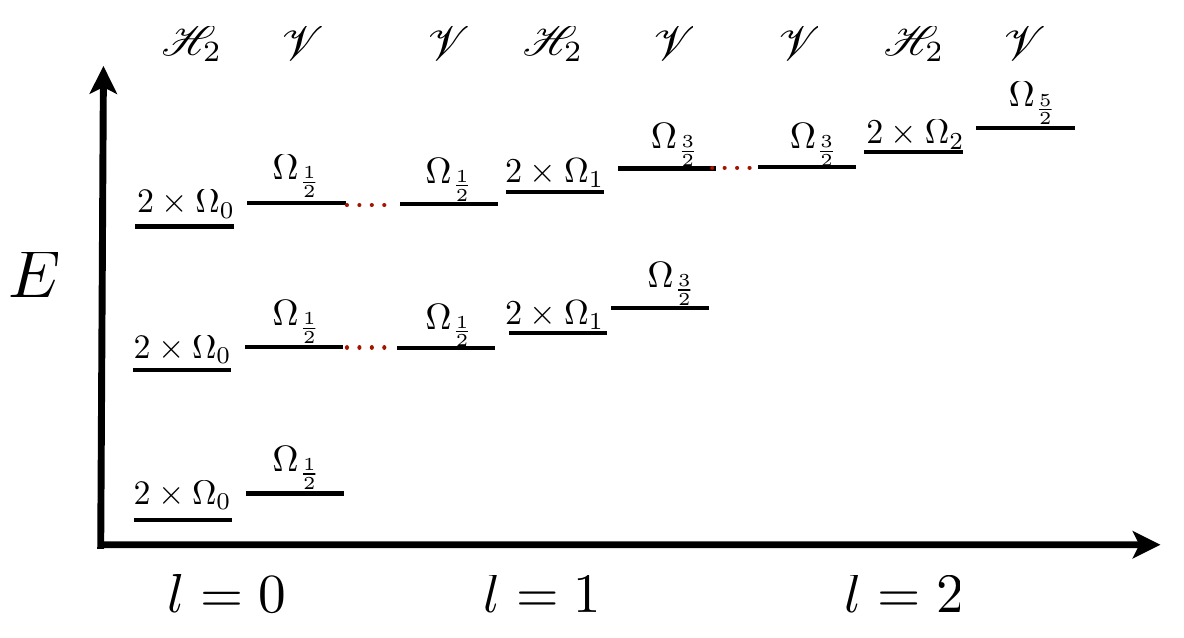}
\caption{\label{Fig: Spectrum} The spectrum of supersymmetric hydrogen up to principal quantum number, $n=3$.  The horizontal
horizontal axis denotes the orbital angular momentum, $l$, beneath while above the superspin state, $|\SS\rangle$,
is denoted above.  The states are labeled by which Clifford they belong to.  The dotted lines show the degeneracies 
that persist after the fine structure is incorporated.}
\end{center}
\end{figure}

Some of the states that do not receive corrections to their energy eigenvalues from second order in perturbation theory are the following states
\begin{eqnarray}
|nlm; l \vec{v}\rangle\qquad |nlm; l\varsigma_-\rangle \qquad |nlm; l \omega_-\rangle 
\end{eqnarray}
where the product between $|nlm\rangle \otimes |\SS\rangle$ is performed to go to the total angular momentum basis.
Only the $j= l$ component of the $|lm\rangle |\vec{v}\rangle$ state doesn't receive a fine structure correction from second order perturbation theory.  
There are no states at second order that can mix these states to $\OO(\alpha^4)$ 
because gaugino exchange has the following selection rules
\begin{eqnarray}
\Delta l =\pm 1\qquad \Delta s = \pm 1 \qquad \Delta j =0
\end{eqnarray}
and the states must be of the same parity.  
These states are representatives of the Clifford vacua
\begin{eqnarray}
\label{Eq: State Ident}
|nlm; l \vec{v}\rangle,\;|nlm; l\varsigma_-\rangle\ \in |n l\rangle \Omega_{l-\half} \oplus |n l \rangle \Omega_{l + \half}, \qquad
|nlm; l \omega_-\rangle  \in |n l\rangle [2\times\Omega_l] .
\end{eqnarray}
Therefore, these states represent elements  of every independent superfield and by computing their energy eigenvalues,
the entire spectrum is computed.  
The perturbing Hamiltonian has non-vanishing matrix elements between these three states and needs to be diagonalized.
Examining the superspin wave functions of these states, only four interactions contribute to $U(\vec{p}_1, \vec{p}_2, \vec{q})$
and are illustrated in Fig. \ref{Fig: Interactions}.    After evaluating the interactions, there is a $3\times 3$ Hamiltonian to diagonalize; however, only $|nlm; l \vec{v}\rangle$ and $|nlm; l\varsigma_-\rangle$ mix.
The interaction matrix is as follows
\begin{eqnarray}
H |\Psi_{nl}\rangle=
 \frac{\alpha^4 \mu}{n^3}\left[ M_n^{\text{FS Univ.}} \identity + M_{nl}^{\text{FS Split}}\right]
 |\Psi_{nl}\rangle
 \end{eqnarray}
where
\begin{eqnarray}
 |\Psi_{nl}\rangle= \left(\begin{array}{c}
|nlm; l \vec{v}\rangle \\
|nlm; l\varsigma_-\rangle\\
|nlm; l \omega_-\rangle 
\end{array}\right) \qquad M_n^{\text{FS Univ.}} = \frac{1}{n} \left( \frac{3}{8} + \frac{1}{8}\frac{\mu}{m_p +m_e}\right)
\end{eqnarray}
and
\begin{eqnarray}
M_{nl}^{\text{FS Split}} = - \frac{1}{2l+1} \identity+\frac{1}{2(2l+1)} \left(\begin{array}{ccc}
-\frac{1}{l(l+1)}&\frac{1}{\sqrt{l(l+1)}} & 0\\
\frac{1}{\sqrt{l(l+1)}}&0 & 0\\
0 & 0 & 0
\end{array}\right) .
\end{eqnarray}
After diagonalizing the interactions and identifying the states with their respective supermultiplets from Eq. \ref{Eq: State Ident},
it is possible to exchange the eigenvalues of orbital angular momentum, $l$, in the expressions for the total angular momentum of the Clifford vacuum,
$j$ ({\em e.g.} $l =j$ for $|nlm; l \omega_-\rangle $).   The mixing angles between $\varsigma_-$ and $\vec{v}$ are precisely the Clebsch-Gordon coefficients for going between the two bases described in Eq.~\ref{Eq: Distinct Wave Functions}. After performing this substitution,
the final fine structure of the supersymmetric hydrogen atom has a remarkably simple (and familiar) expression given by
\begin{eqnarray}
\label{Eq: Spectrum}
E_{nj}= E_n^{\text{Ryd}} + \Delta E_n^{\text{FS Univ.}} + \Delta E_{nj}^{\text{FS Split}}
\end{eqnarray}
where the leading order bound state energy is $ E_n^{\text{Ryd}}=-\alpha^2\mu/2n^2$.   The fine structure is divided into a $j$-independent
universal term and a term that leads to a fine structure splitting
\begin{eqnarray}
\Delta E_n^{\text{FS Univ.}}=\frac{\alpha^4\mu }{n^4}\left(\frac{3}{8}-\frac{\mu }{8(m_p+m_e)}\right)
\qquad
\Delta E_{nj}^{\text{FS Split}}=-\frac{\alpha^4\mu }{2n^3}\frac{1}{j+\half} .
\end{eqnarray}
The spectrum is schematically illustrated in Fig. \ref{Fig: Spectrum}.
As expected, the fine structure reduces to the non-supersymmetric limit of Eq. \ref{Eq: HydrogenSpectrum} 
when \mbox{$m_e/m_p \rightarrow 0$.}  The next feature is that
the spectrum is independent of the orbital angular momentum, $l$, as is the case in the  hydrogen atom.  The last pertinent feature
is that the $j$-dependent,  fine structure splitting receives no supersymmetric contributions.

\subsection{Supersymmetric Lamb Shift}
\label{Sec: Lamb}

The degeneracies of the supersymmetric hydrogen atom at principle quantum number $n$ are $2n-1$ two-fold degeneracies.
The two-fold degeneracies for the states that are built from integer spin Clifford vacua arise from the $O(2)_R$ symmetry and unless parity or the $U(1)_R$ is broken by some other interactions, these degeneracies will persist to all orders in the
relativistic expansion as well as to all orders in perturbation theory.  

The two-fold degeneracies for the half-integer Clifford vacua is an accident and it is natural to ask at when is the accidental degeneracy lifted.   This degeneracy is completely analogous to the corresponding degeneracy for the Dirac hydrogen atom
and it is well known that the Lamb shift arises radiatively.  In this section, the correction from the Lamb shift is estimated.
This article only addresses the logarithmically enhanced correction in the infinite proton mass limit. 
The infinite proton mass limit is particularly illuminating because the accidental degeneracy occurs for supermultiplets
that have valence fermionic electrons.  The logarithmically enhanced contribution traces back to the collinear singularity
of the photon and electron and doesn't exist for scalars external legs or photino interactions.  Additionally there are fewer interactions because all supersymmetric interactions drop out as discussed in Sec. \ref{Sec: HeavyP}.
The finite correction and corrections away from the infinite proton mass are left for further study.

In  QED there are two  one-loop processes contributing to the Lamb shift and the IR behavior of the integral is
challenging due to the collinear and soft divergences in QED.  These divergences are regulated by the discreteness of the hydrogen spectrum.  $\Lambda_{\text{IR}}$ is the scale where the IR divergences are cut-off.
    Parametrically $\Lambda_{\text{IR}} \sim \alpha^2 m_e$ and detailed calculations give
$\Lambda_{\text{IR}} \simeq 10 \alpha^2 m_e$.

\begin{figure}[htbp]
\begin{center}
\includegraphics[width=3.5in]{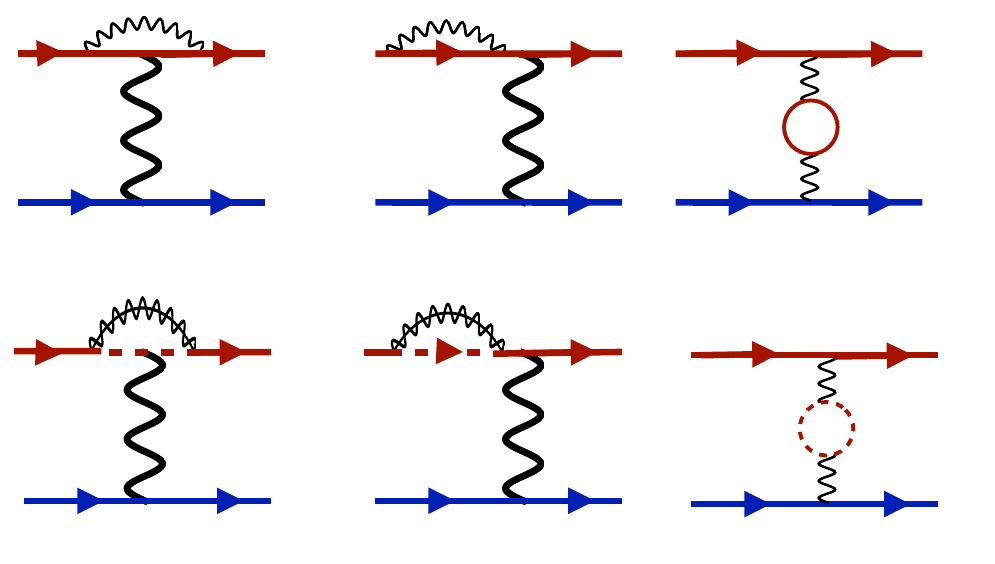}
\caption{\label{Fig: Lamb} The SQED contributions to the Lamb shift.  Only the upper left diagram has
a logarithmically enhanced contribution to the splitting.  This contribution dominates and therefore the
dominant contribution to the Lamb shift is identical between QED and SQED.}
\end{center}
\end{figure}

All corrections from gaugino exchange and contact interactions from $D$-terms vanish in the non-relativistic limit as $m_p\rightarrow \infty$.  The renormalization of supersymmetric exchange diagrams can not be parametrically enhanced.  The self energy contribution from either the photon or photino do not have IR-divergences because the virtual particles in the loop are massive.  The gaugino corrections to the vertex do not have IR-divergences. Therefore, the supersymmetric contributions to the logarithmically enhanced portion of the Lamb shift
reduce to the QED contributions.
Therefore the dominant contribution to the Lamb shift is given by \cite{LandauLifschitz}
\begin{eqnarray}
\Delta E_{nl} \simeq  \frac{ 4 \alpha^5 m_e}{3 \pi n^3} \log \frac{ m_e}{\alpha^2 m_e} \delta_{l 0}  + \cdots.
\end{eqnarray}
The terms that are not logarithmically enhanced  will not have the same universal form and need to be computed.  

\section{Non-Abelian Bound States}
\label{Sec: Non-Abelian Bound States}

Hydrogen is a useful model for many other bound states and provides intuition about more complicated systems.
For instance, the constituent picture of mesons and baryons is heavily based on Coulombic bound states of quarks
as the fine structure constant goes towards unity.    This section lays out how the formalism above is used for
other bound state systems.

The most straightforward extension is to promote the binding $U(1)$ to an $SU(N_c)$ gauge theory with the electron and
proton becoming fundamentals and anti-fundamentals of the gauge group.  When the confinement scale
$\Lambda \ll m_e, m_p$, the electron-proton bound states should be approximately Coulombic.   The effective coupling between
the proton and electron is now given by 
\begin{eqnarray}
\alpha(\mu) = C_2 \frac{g^2(\mu)}{4\pi} .
\end{eqnarray}
The primary difference between the SQED bound states and SQCD bound states is that the gauge coupling runs.  The running of the gauge coupling breaks the accidental $O(4)$ symmetry
of the Coulomb problem.   Without the $O(4)$ symmetry, the $\OO(\alpha^2 \mu)$ energy eigenvalues are
depend upon the orbital angular momentum.   The bound state spectrum of this theory still has a multi-fold degeneracy arising from  different superspin configurations  ($\Omega_{l-\half}$, $\Omega_l$, and $\Omega_{l+\half}$) for a given orbital angular momentum, $l$.  This remaining degeneracy is lifted by the fine structure.   The non-Coulombic nature of the running coupling constant will make evaluating fine structure through second order perturbation theory difficult.  However, the method of identifying the states that do not have corrections to their energy eigenvalue from second order perturbation
theory is still applicable.

A related example to the previous case is to keep the proton and electron heavy 
and add in $N_f$ light spectator quarks that slow the running of $\alpha(\mu)$.
In the limit $N_f\simeq 3N_c$ the theory enters a weakly coupled fixed point in the IR and the leading order potential becomes approximately Coulombic and the $O(4)$ symmetry re-emerges.   The arguments from Sec. \ref{Sec: NR} indicate that the fine structure degeneracy will emerge in the 
infinite proton mass limit.  The Lamb shift is more subtle in this case because the gluons of the $SU(N_c)$ gauge theory are colored.  The logarithmically enhanced contribution is related to emitting a gluon; however emitting a gluon  from
the binding, color-singlet state will transition the bound state into a color-adjoint configuration that doesn't bind.   The virtual gluon has energy of the order of a Rydberg and the energy-time uncertainty softens this effect.  Nevertheless, this discussion indicates that the Lamb  degeneracy is potentially more interesting for these non-Abelian gauge theories. 

\subsection{Heavy-Light Bound States}
\label{Sec: Heavy-Light}

The final example of a non-Abelian bound state that this article considers is a more subtle situation when there is
one light quark (the electron) and one heavy quark (the proton), relative to the scale of confinement.    Non-Abelian super-Yang Mills theories with $N_F  \le N_c$
do not have a vacuum in the absence of supersymmetry breaking.  Instead, the light quarks develop an Affleck-Dine-Seiberg superpotential
that is singular at the origin
\begin{eqnarray}
W_{\text{ADS}}\simeq  \Lambda^3 \left(\frac{\Lambda^2}{E E^c}\right)^{\frac{1}{N_c}}
\end{eqnarray}
where $E$ and $E^c$ are the light-flavored quarks. 
In the presence of soft supersymmetry breaking, soft masses for the squarks will stabilize the run-away potential.
The location of the vacuum is roughly
\begin{eqnarray}
\langle \tilde{e}\rangle = \langle\tilde{e}^c\rangle \sim\Lambda \left(\frac{\Lambda}{m_{\text{soft}}}\right)^{\frac{N_c}{2 (N_c+1)}}  \simeq m_V.
\end{eqnarray}
These scalars can only break $SU(N_c)\rightarrow SU(N_c-1)$ and it gives mass to vector superfields transforming as $\square + \overline{\square} +\mathbf{1}$ under the $SU(N_c-1)$.
The electrons get eaten by the super-Higgs mechanism and absorbed into the massive vectors.
This stabilizes the gives mass to the vectors, $m_V$, above the confinement scale so long as $\Lambda \gsim m_{\text{soft}}$.
In this is the case, the massive vectors are quasi-weakly coupled and the resulting spectrum of the theory is qualitatively different than the those of supersymmetric hydrogen.

The heavy proton, after the scalar electron acquires a vacuum expectation value, decomposes into 
\begin{eqnarray}
P \rightarrow \mathbf{1} +\square, \qquad P^c \rightarrow \mathbf{1} + \overline{\square}.
\end{eqnarray}  
Since the $SU(N_c-1)$ has no light flavors, it confines at low energies and the colored proton must be screened.
The proton can only be screened by an anti-proton or one of the massive vectors (that have eaten the electrons).
Because the vectors are heavier than the confinement scale, the proton-vector bound states are quasi-Coulombically bound.
The procedure from Sec. \ref{Sec: Sym} can be performed again, but with $|\SS\rangle$ now given by the product
of a hypermultiplet and a massive vector field:
\begin{eqnarray}
|\SS\rangle_{\vec{p}=\vec{p}_{\text{cm}}} =  (P \oplus P^c{}^\dagger)_{\vec{p}=\vec{p}_{\text{cm}}+ \vec{p}_{\text{rel}}/2} \otimes V_{\vec{p}=\vec{p}_{\text{cm}}- \vec{p}_{\text{rel}}/2}
\end{eqnarray}
This product can be decomposed into
\begin{eqnarray}
|\SS\rangle = \begin{cases}
\RRR \equiv \Omega_1 & \mbox{massive, complex spin $\frac{3}{2}$ supermultiplet}\\
 \VVV_2 \equiv 2\times \Omega_\half & \mbox{massive, complex vector supermultiplet}\\
\HHH \equiv \Omega_0 & \mbox{massive, complex hypermultiplet}
\end{cases}
\end{eqnarray}
The $O(2)_R$ symmetry is still present in the theory and causes 
the doubling of the vector supermultiplet.
Since the ground state will  have no orbital angular momentum, $|\SS\rangle$ are equivalent to the ground state degeneracy
of the system.  The fine structure will split these multiplets.  

In addition to the supersymmetric corrections to the spectrum, supersymmetry breaking effects arise from the soft masses of the theory;
however, these can be made arbitrarily small in the limit $m_{\text{soft}}/ \Lambda\rightarrow$ holding $m_V/m_p$ fixed. 

\section{Discussion}
\label{Sec: Conc}

This article calculated the spectrum of supersymmetric hydrogen and yielded a spectrum that is nearly identical to that of the Klein-Gordon and Dirac equation even for finite proton masses.   The surprisingly mild effect of the plethora of supersymmetric interactions is closely related to the non-relativistic limit.  The fine structure degeneracy is still present in the supersymmetric hydrogen atom and this degeneracy is lifted by the Lamb shift.  The logarithmically enhanced contribution to the Lamb shift is the same as in QED, but the finite contributions are sensitive to supersymmetric interactions.  The Johnson-Lippmann operator guarantees the $l$ independence of the energy eigenvalues in the infinite proton mass limit for both supersymmetric and non-supersymmetric QED.  For a finite proton mass in non-supersymmetric QED there is no degeneracies of the fine structure and therefore no Johnshon-Lippmann operator.  However, in supersymmetric QED, the residual degeneracy of the fine structure indicates that  a generalization of the Johnson-Lippmann operator should exist even with a finite proton mass and this operator has not been found.  The methods presented in this article are widely applicable to other supersymmetric bound states and additional applications were briefly discussed in Sec. \ref{Sec: Non-Abelian Bound States}.   
 
 In addition to studying the spectroscopy of atoms, supersymmetry also constrains the interactions of the bound states.  The interactions of the ground state with an external fields (such as the vector superfield) should be described in terms of an effective field theory.    Studying the interactions may giver further insight into the dynamics of the bound states. Ultimately, some variant of these bound states might be dark matter.  The astrophysical anomalies currently   observed  hint that dark matter may have more structure than typically assumed, see \cite{ArkaniHamed:2008qn} for a recent description.   The tools presented in this article may
help construct models to further explore these possibilities.   The theory studied in Sec. \ref{Sec: Heavy-Light} is the supersymmetric generalization of composite inelastic dark matter \cite{Alves:2009nf, Lisanti:2009am}.
The discussion in Sec. \ref{Sec: Heavy-Light} indicates that there is a much richer structure than the simple generalization of mesons to supersymmetric mesons because of the vacuum structure of the theory.

If this supersymmetric gauge theory is coupled to the Standard Model, supersymmetry breaking
will be communicated to the spectrum at some level.  At this stage, it is not possible to cherry pick 
convenient states to compute the effects of supersymmetry breaking; however, assuming
that supersymmetry breaking is a modest effect, it is only necessary to use first
order in the supersymmetry breaking interactions.  The leading order splitting is induced by the scalar
soft masses.  This perturbation is straight-forward.  The more challenging measure is the two-fold degeneracy of states in the $|\SS\rangle =\HHH_2$ 
configuration will only be lifted through parity breaking or $R$-breaking.  These effects are model
dependent and a specific example is considered in \cite{Upcoming}.
 
\section*{Acknowledgements}
We thank S. Behbahani for extensive collaboration at the earliest stages of the work.
We would like to thank M. Jankowiak for many insightful conversations during
this work including insights into the $O(2)_R$ symmetry.  We would also like to thank  D. Alves for verifying many of the calculations contained in this draft.  JGW would like to thank
J. Kaplan for helpful discussions on on-shell supersymmetry and the Lamb shift.
JGW thanks E. Katz for useful conversations during the course of this work.

TR is a  William R. and Sara Hart Kimball Stanford Graduate Fellow.
  JGW  is supported by the US DOE under contract number DE-AC02-76SF00515.  
TR and JGW receive partial support from the Stanford Institute for Theoretical Physics.  
JGW is partially supported by the US DOE's Outstanding Junior Investigator Award.  

After this work was completed, \cite{Herzog:2009fw} appeared and initially had minor
discrepancies with the results in this paper.   We thank C. Herzog examining our draft
before publication and finding a typo in a critical expression.  

\providecommand{\href}[2]{#2}\begingroup\raggedright

\endgroup

\end{document}